\documentclass[twocolumn,epj]{webofc}
\usepackage[varg]{txfonts}   
\usepackage{graphicx}  
\usepackage{dcolumn}   
\usepackage{bm}        
\usepackage{amssymb}   
\usepackage{amsmath}
\usepackage{xspace}
\usepackage[caption = false]{subfig}

\newcommand{\be}{\begin{equation}}
\newcommand{\ee}{\end{equation}}

\newcommand{\xmax}{$X_{\rm max}$}

\newcommand{\mxm}{$\langle X_{\rm max} \rangle$}
\newcommand{\rmsx}{${\rm RMS}[{X_{\rm max}}]$}

\newcommand{\sigcl}{$\sigma_{\rm cl}$}

\woctitle{?????????}

\begin{document}

\title{Particle Physics at Ultrahigh Energies
}

\author{Glennys R. Farrar }
\institute{Center for Cosmology and Particle Physics,
Department of Physics,
New York University, NY, USA}
\date{\today}

\abstract{
We explore particle physics beyond accelerator energies, motivated by questions exposed in astroparticle physics observations:  1) Are there reasonable modifications to the standard extrapolations of LHC-tuned hadronic interaction models, so that ultrahigh energy cosmic ray (UHECR) showers are well-described with a purely protonic primary composition rather than requiring a tuned, energy-dependent composition mixture as needed in conventional models?  2) What modifications to standard models can solve the deficiency in the predicted ground signal found in hybrid UHECR observations?   We find that a pure proton composition provides an excellent fit to shower observations, if the QCD inelastic cross section increases more rapidly above $E_{\rm cm} \approx 60$ TeV than in conventional models, and speculate as to possible reasons this may happen; the ``muon deficiency'' can be cured by relatively minor modifications to particle ratios in unexplored kinematic regimes below and above LHC energies.  (Note added:  This paper was prepared for the ISVHECRI 2014 proceedings, but never posted to the arXiv through an oversight.  A few relevant subsequent citations have been added.)
}
\maketitle 

\section{Introduction}

A crucial question for unraveling the puzzle of the nature of Ultrahigh Energy Cosmic Rays is to know whether UHECRs are predominantly protons or a mixture of different nuclei.  However inferring the composition from the properties of the UHECR air showers requires extrapolating models of particle physics into kinematic regimes and energies not accessible at accelerators.   For instance, the nucleon-nucleon center-of-mass energy $E_{CM}$ in the first interaction of a 10 EeV ($10^{19}$ eV) UHECR is 137 TeV,  and tens to hundreds of the secondary interactions are above the LHC energy.  Adding concern that the particle physics involved is inadequately understood, is the fact that even though the leading models have been tuned to fit the latest accelerator data, their predictions appear not to consistently describe both the shower development in the atmosphere and the signal in ground detectors  \cite{topdownPRL16}.

Extrapolations of QCD modeling can fail for several reasons, the most interesting of which are a transition in the dynamics of QCD, e.g., due to gluon saturation, or to the opening up of new interactions.  In this note we develop several scenarios of such transitions to identify examples of phenomena which may be occuring in UHE air-showers.  

\section{\xmax\ in UHECR Air Showers}

The primary tool for inferring composition is the distribution of the depth in the atmosphere at which individual UHECR showers reach their maxima, the \xmax\ distribution.  For events detected with the air fluorescence detector telescope (FD), the longitudinal development of the atmospheric shower can be measured.   The integrated FD signal provides the energy of the primary UHECR, and the column-depth at which the individual shower peaks gives \xmax.  The value of \xmax\ for a given shower is determined by the depth of first interaction and also how rapidly the shower develops, as discussed in \cite{KampertUnger12} and references therein.  The effective depth of first interaction is governed by the inelastic scattering length $\lambda_{\rm inel}$, and thus its distribution is deeper and broader for a proton primary than for nuclei.  Furthermore,  the shower from a nuclear primary is more homogeneous and builds up more quickly than for a proton primary since roughly speaking it consists of $A$ superimposed sub-showers, each of energy $E/A$.  

\begin{figure}[t!]
\centering
\includegraphics [width=\columnwidth]{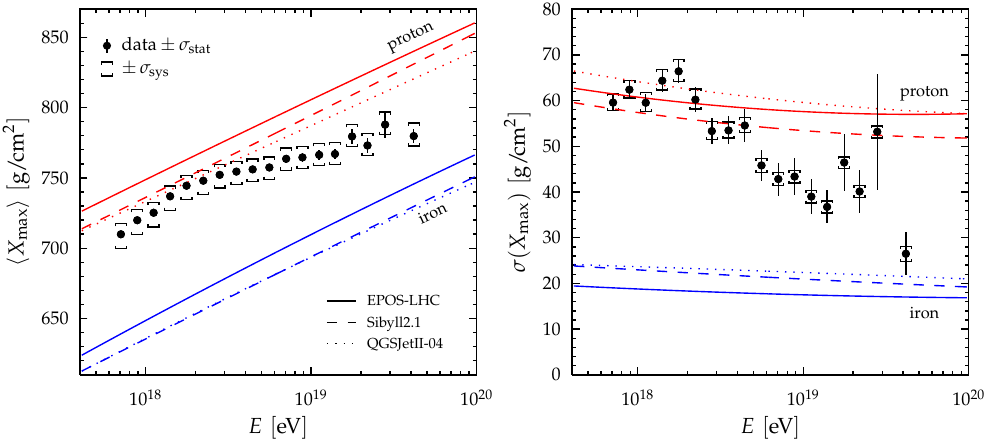}
\caption{Energy dependence of  \mxm\  and \rmsx\ from \cite{augerXmaxMeas14}, compared to predictions of various HEGs for p and Fe primaries.}
\label{ElongRate}
\end{figure}

Fig. \ref{ElongRate} shows the evolution in energy of \mxm\ (the ``elongation rate'') and of the RMS width of the \xmax\ distribution, from Auger \cite{augerXmaxMeas14}.  The predictions for pure p and pure Fe composition in the current generation of hadronic event generators (HEGs) -- QGSJet-II-04 \cite{QII-04} and EPOS-LHC \cite{EPOS-LHC}, which have been re-tuned to LHC data, and a third model Sybill \cite{Sibyll09} -- are shown by the red and blue lines, as indicated.  Note that Telescope Array reports a composition consistent with pure proton, but the TA data are also consistent with Auger's results when systematic and statistical uncertainties are taken into account \cite{augerTA_wgComp15}.

The conventional interpretation of these observations is that the composition is becoming heavier with increasing energy.  It is important to emphasize that the data is incompatible with a proton-Fe mix with gradually increasing Fe fraction:  that would lead to an \rmsx\ which at the 50-50 point is greater than the difference between the means of the p and Fe \mxm\ values;  thus in a p-Fe scenario, \rmsx\ should increase from the protonic value of about 60 g cm$^{-2}$ to  $\gtrsim$100 g cm$^{-2}$, before dropping to about 25 g cm$^{-2}$ when the composition is nearly pure Fe.  Data instead shows a steadily decreasing \rmsx.  In order to interpret the \rmsx\ behavior as a changing composition, the composition has to change gradually, moving first through light, then intermediate, then heavier masses.  This is puzzling astrophysically, because light and intermediate mass nuclei break up from photodisintegration, ultimately producing protons of energy $E/A$, over much shorter propagation distances than do heavy nuclei and on a distance scale short compared to the energy loss length for protons due to photopion and $e^{\pm}$ production.  Thus even if light and intermediate masses are present at the source in comparable abundances to protons and heavy nuclei, it requires a very hard source spectrum and  finely tuned composition to reproduce the observed \xmax\ distribution and its energy dependence. \footnote{Since this work presented at ISCHECRI 2012 , it has been shown that the spectrum, mixed composition and its evolution, and the light component below the ankle, are well-explained if an intermediate mass nucleus, or a composition mix such as seen in Galactic cosmic rays, pass through a photon field surrounding the accelerator, which somewhat fragments the nuclei before their journey to Earth \cite{ufa15}, diminishing the need for an alternative to the ad-hoc mixed composition models which failed to explain the ankle shape and the light but extragalactic UHECR population below the ankle.  Nonetheless, it is interesting to see how new physics can reproduce much of the observations, with pure protonic UHECRs.} 

\begin{figure}[t!]
\centering
\includegraphics [width=0.48\columnwidth]{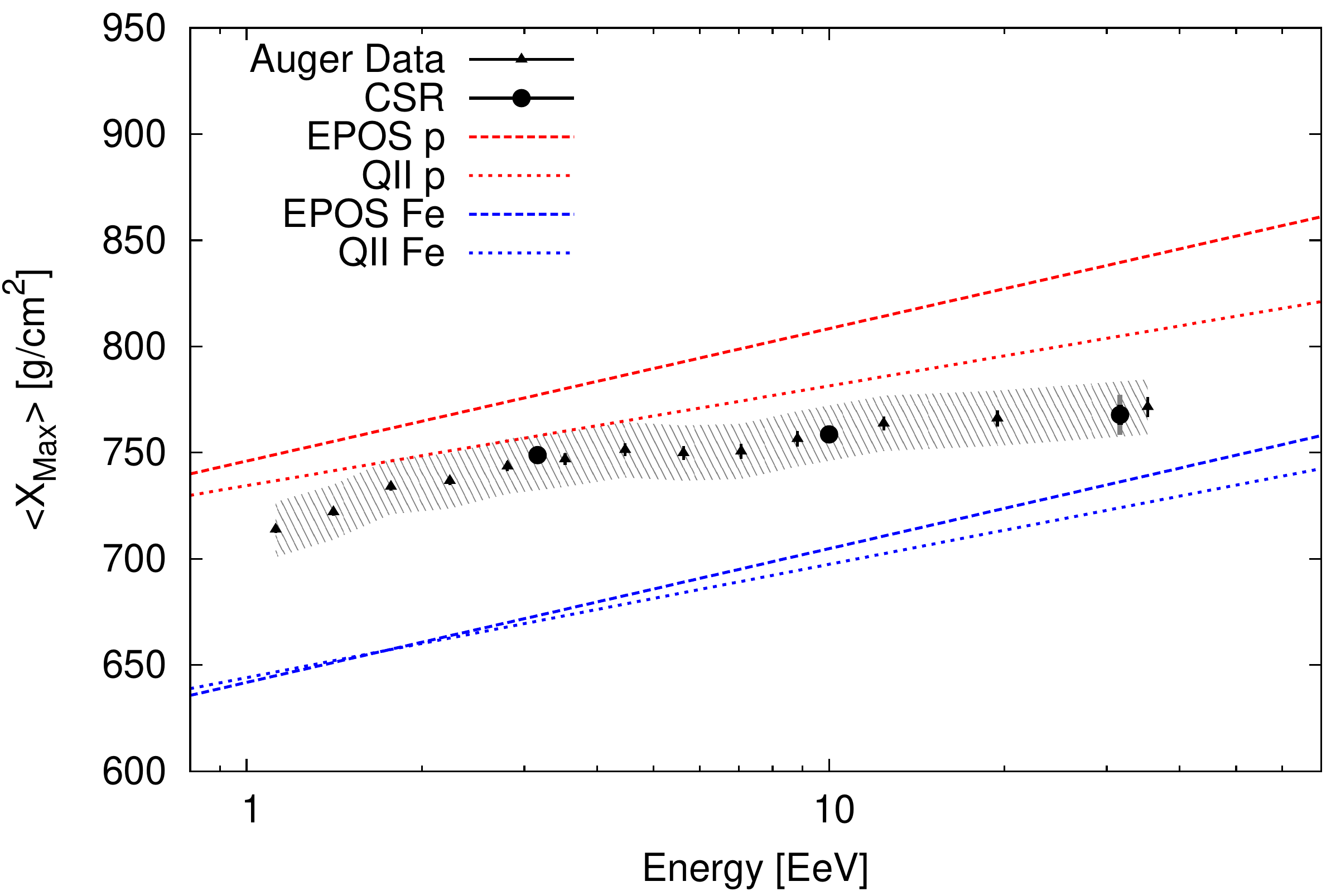}
\includegraphics [width=0.48\columnwidth]{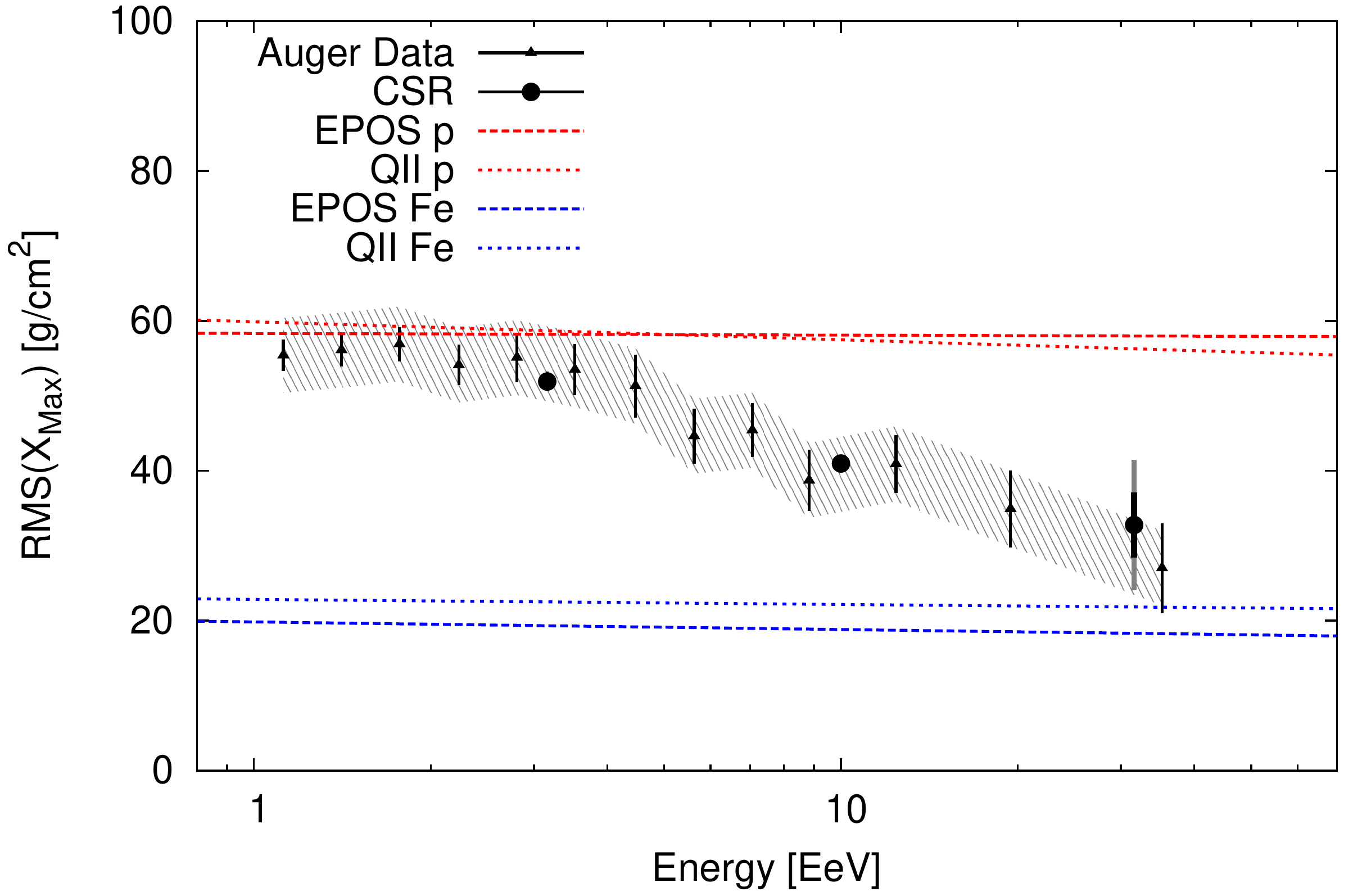}
\caption{Energy dependence of  \mxm\  (shifted by 5 g cm$^{-2}$, which is about half the 11 g cm$^{-2}$ systematic uncertainty quoted by Auger) and \rmsx\ , compared to predictions of model with increased cross section, labeled CSR.}
\label{xmaxCSR}
\end{figure}

Without entering into the debate about the astrophysical plausibility of mixed composition scenarios, here we explore an alternate interpretation of the \xmax\ observations.  We show that a good accounting of the energy evolution of \mxm\ can be obtained with a purely protonic composition, if the growth in the inelastic p-Air cross section assumed in the standard HEGs is modified.   Remarkably, a pure proton composition with increasing cross section, which fits the elongation rate, also gives an excellent fit to the observed behavior of \rmsx\ and vice-versa.  This fit is shown in the solid black curves in  Fig. \ref{xmaxCSR}.   Reinforcing the case for this interpretation, is the fact that it gives a good description of the {\em full} \xmax\ distribution -- not just its mean and RMS -- {\em over the entire observed energy range}.  

\begin{figure}[t]
\centering
\includegraphics [width=\columnwidth]{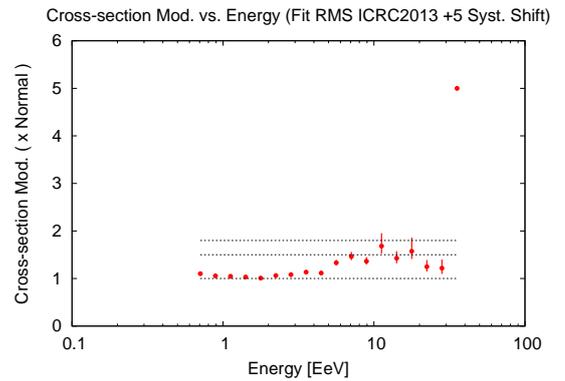}
\caption{The factor by which the cross section assumed in a typical HEG, that of QGSjetII-04, needs to be increased to fit the observed behavior of the \xmax\ distribution.}
\label{modsig}
\end{figure}

\begin{figure}[t]
\centering
\includegraphics [width=\columnwidth]{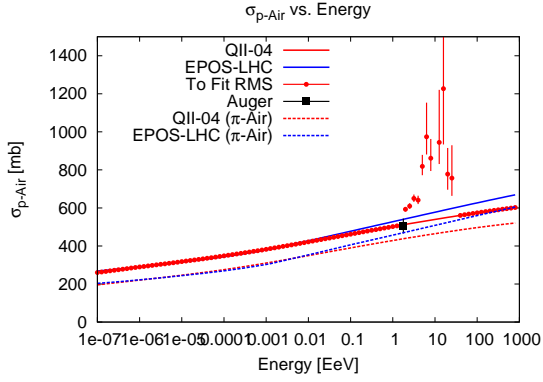}
\caption{The energy dependence of the p-Air cross section assumed in various HEGs, compared to that used for the pure-proton scenario; the data point is the Auger measurement \cite{augerXsecnPRL} .}
\label{sigvsE}
\end{figure}

\begin{figure}[t!]
\centering
\includegraphics [width=\columnwidth]{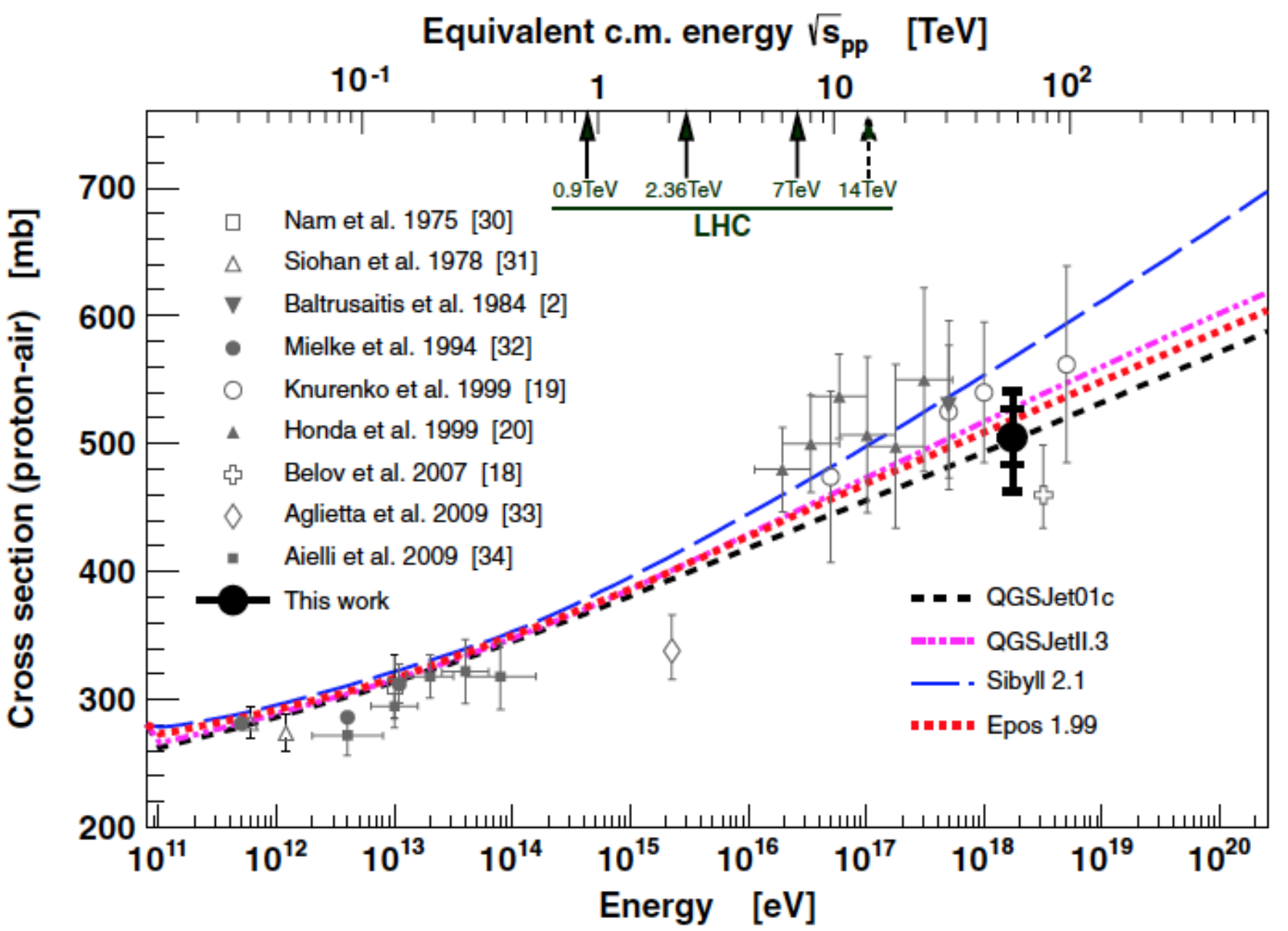}
\caption{The p-Air cross section as a function of energy, from \cite{augerXsecnPRL} .}
\label{sigpAir}
\end{figure}

In this fit displayed in Fig. \ref{xmaxCSR}, we have used QGSJetII-04 as the underlying hadronic event generator (HEG) but increased the p-p cross section relative to the default value by the factor shown in Fig. \ref{modsig}.  For orientation, the behavior of the measured cross sections and standard extrapolations are shown in Figs. \ref{sigvsE} and \ref{sigpAir}.  The modification is suggestive of a threshold rather than being gradual, but is also compatible with a smoother increase such as in the ``QCD classicalization'' scenario discussed below, given the statistical and systematic uncertainties of the data.

In the remainder of this note we explore mechanisms which could give rise to the modified behavior of the cross section shown in Fig. \ref{modsig}.  

\section{Gluon Saturation}
It is well-known in the context of relativistic heavy-ion collisions that, at sufficiently high energy, the density of partons becomes so large that partons can no longer be considered  weakly-interacting, and non-linear effects such as gluon saturation and recombination become important.  Qualitatively, partons (which are predominantly gluons) recombine and diffuse transversely in order to moderate the growth in density.  It is instructive to compare the surface  energy density -- total column energy per effective nucleon transverse area --  in three situations: Pb-Pb and p-Pb collisions at the LHC, and UHE p-Air collisions.   Treating a nucleus of mass $A$ as a cube of side $A^{1/3}$ gives the surface energy density in GeV/nucleon-cross-section listed in the table

\begin{center} \begin{tabular}[h]{lll}
initial state & mean column-energy & in GeV \\
\hline
Pb+Pb(1.38TeV per N) & $208^{\frac{1}{3}} \times 2.76$ TeV & $1.7 \,10^{4}$ \\
p(3.5TeV)+Pb(1.38TeV)  & $(208^{\frac{1}{3}} \times 1.38 + 3.5) $TeV & $1.2 \, 10^{4}$ \\
p($10^{18.5}$eV)+A($m_{p}$) & $(2 A m_{p} E_{\rm CR})^{\frac{1}{2}} (1 + A^{-\frac{2}{3}})$ & $1.7 \, 10^{5} $ \\
\end{tabular} \end{center}

\noindent Even though these estimates are simplistic, they show that p-Air collisions at $10^{18.5}$ eV, where the slope of the Elongation Rate changes, entail column-energies a factor 10 higher than ever encountered in the laboratory.

Some consequences of the gluon saturation phenomenon are already seen in LHC heavy-ion collisions; see \cite{amGluonSatReview14} for a recent review.  The basic idea is that when the density of gluons gets big enough, they cannot be treated perturbatively\footnote{The event generator EPOS makes an effort to model gluon recombination, but in the absence of better developed non-perturbative techniques, this is necessarily an empirical effort aimed at reproducing observations at the LHC.}.  At a minimum, gluon recombination must be considered in order to preserve unitarity, but additional effects can be profound.  The space-time picture of a stationary parton cloud is clearly insufficient.  The system can display coherent, long-range interactions which force partons to larger transverse distance (thereby increasing the total cross section) and may simultaneously lead to larger final state multiplicities.  

The main message we take away is that a rapid, O(1) increase in cross section in p-Air collisions above some critical parton density is intuitively plausible, and that the energy scale at which this could produce a break in the elongation rate is not unreasonable.  A natural corollary is an increase in multiplicity, which can contribute both to more rapid development of the shower and an increase in the muon signal at ground.   

\section{Classicalization}

As part of an effort to understand the hierarchy between the electroweak scale and the Planck mass, Dvali and collaborators \cite{dvali+classicalization11} proposed that under certain conditions a theory can exhibit a phenomenon called classicalization.  Dvali and Gomez  \cite{dgUHEclass12} explored possible signatures of classicalization of gravity in UHE particle scattering and other laboratory settings.  In the application to general relativity, the elementary quanta are gravitons which they suggest may form collective modes (or many modes with a tower of masses), characterized by a mass-scale $\mu$ of order the electroweak scale.  Here we apply their classicalization analysis to QCD in the regime of extreme gluonic densities.

The physical picture is that at high densities the appropriate degrees of freedom change from being individual partons to collective states of relatively weakly interacting bosons, which may have spin-0, spin-2 etc.  On very general grounds, the leading high-energy interaction of the spin-two degree of freedom $h_{\mu \nu}$ is
\be
\label{hmunuTmunu}
{\cal L}_{\rm eff} \equiv \frac{1}{M*} h_{\mu \nu} \, T^{\mu \nu},
\ee
where $T^{\mu \nu}$ is the energy-momentum tensor; the scale $M*$ is related non-perturbatively to the deBroglie wavelength $\mu$ of the collective mode $h_{\mu \nu}$:  
\be
\label{M*mu}
M* = \mu \, e^{1/g}.
\ee
Spin-0 degrees of freedom in general have an analogous coupling to the energy momentum tensor as in Eq. \ref{hmunuTmunu}, with $h_{\mu \nu} \rightarrow \partial_{\mu} \partial_{\nu} \phi$.  In the remainder of this note we will simply adopt the coupling Eq. \ref{hmunuTmunu} since the collective modes formed from gluons should include spin-2 modes, for which the UHE scattering can be directly analogized to the discussion of \cite{dgUHEclass12}.  I.e., we replace $M_{\rm Planck} \rightarrow M*$, and the electroweak scale $\rightarrow \mu$, with $\mu$ expected to be of order $\lesssim100$ MeV in QCD.

At leading order, the relation between the radius of the classicalized state in terms of the CM energy $E = \sqrt{s}$ and the scale $M*$, is the same as the relationship between the Schwarzschild radius of a black hole and its mass: 
\be
\label{Rcl}
R_{\rm cl}(E) = \frac{E}{M*^{2}}~~  \leftrightarrow ~~R_{\rm Sch}  = \frac{M_{\rm BH}}{M_{\rm Planck}^{2}}. 
\ee
The occupation number, $N$, of collective modes in a system of total energy $E$ and size $\lambda$ is $N \approx E \lambda $. In the classicalized state of energy $E$, the characteristic wavelength of the collective modes $\lambda$ $ \approx R_{\rm cl}(E)$ increases with energy until such high energies that $R_{\rm cl}(E) \sim \mu^{-1}$; at higher energies corrections reduce the growth of $R_{\rm cl}$ with $E$.  Taking into account these corrections, \cite{dgUHEclass12} finds
\begin{subequations}
\begin{align} 
\label{sigcl}
\sigma_{\rm cl} &= \frac{\pi \, s}{M*^{4} }{\rm Exp} [ \,2\, \mu \, \sqrt{\sigma_{\rm cl} / \pi }\, ]    
\\ & \label{sigclb} \rightarrow \frac{2\, \pi \,m_{p} \, E_{\rm CR}}{M*^4},                
\end{align}
\end{subequations}
with the limiting second form, Eq. (\ref{sigclb}), applicable in the threshold region and the more general behavior determined by the trancendental equation Eq. (\ref{sigcl}) guaranteeing asymptotic unitarity.  Due to the rapid growth in the classicalization cross section in the threshold region, if $M*$ is small enough for classicalization to occur in UHECR collisions it will have a significant impact on the \xmax\ distribution.  One can fit the required enhancement shown in Fig. \ref{} with $M* \approx 200$ GeV.  

The second relevant effect of classicalization on UHECR air showers arises because the final states produced by the decay of the classicalized configuration can be expected to differ from final states in conventional partonic interactions in two ways:   
\begin{itemize}
\item Generally larger multiplicity  
\item Particle production mediated by the decay of the collective modes, leading to final states enriched by the decay products of glueballs.
\end{itemize}
We can estimate the number of final particles resulting from the decay of a classicalized system from $N \approx E R_{\rm cl}(E) $, with $ R_{\rm cl}(E) $ given by Eq. \ref{Rcl}:
\be
\label{Nf}
N_{f}(E) = \kappa \, \frac{s}{M*^{2}}  \left(\frac{\mu}{M_{\rm gb}} \right),
\ee
where the glueball mass scale $M_{\rm gb} \approx 1.5-2$ GeV and $\kappa$ absorbs the uncertainty in the hadronization process of the classicalized configuration but should be of order 0.1-1. 

Chanowitz \cite{chanowitz05} argued that $K-\bar{K}$ final states should dominate over $\pi-\pi$ final states in the decay of spin-0 glueballs because of chirality constraints.  He identified the $f_{0}(1710)$ as the groundstate glueball and indeed the ratio of branching fractions for BR($\pi-\pi$)/BR($K-\bar{K}) < 0.11$ at 95\% CL in  $f_{0}(1710)$  decay.  Following this reasoning, we can expect that the K/$\pi$ ratio in final states from classicalization will be enhanced relative to that in conventional HEGs.  Since at this time we cannot estimate what fraction of final particles in the decay of the classicalized configuration arise from decay of spin-0 glueballs, we will take the classicalization final states to have a $K/\pi$ ratio $r_{K}$ .   An enhanced $K/\pi$ ratio in classicalization final states brings the predicted and observed ground signal in hybrid showers into better agreement\cite{afICRC13}.

\section{Phenomenological Implementation of QCD-Classicalization for UHE Air Showers}

The classicalization picture has the potential to provide a good and very economical fit to the measured elongation rate and its RMS width.  Furthermore, it can address a persistent problem in modeling UHECR showers -- that the strength of the hadronic shower at ground level in hybrid events at $10^{19}$ eV is at least a factor 1.3 too low using the LHC-tuned HEGs \cite{topdownPRL16}.  It may also improve the relationship between the muon production depth and \xmax\ distributions.   In this section, we outline an approach to fitting the UHECR shower data, with 4 free parameters of the phenomenological classicalization model outlined above.   

We assume that the primary CR is a proton, and that the relation between p-Air and p-p interactions is adequately treated by Glauber theory as implemented in standard hadronic event generators (HEGs).  In the following, the specific procedure used in \cite{afICRC13} to model new physics is described; it can clearly be adapted to using a different, newer HEG.

A very important conceptual point is that because the classicalization final states are highly dissimilar with respect to the final states in conventional collisions, the conventional and classicalization contributions to the scattering amplitude do not interfere and the total scattering probability is the incoherent sum of the two contributions:
\be
\label{sigtot}
\sigma_{\rm tot} = \sigma_{\rm HEG} + \sigma_{\rm cl},
\ee
with \sigcl\ determined from Eq. \ref{sigcl}.    

We take QGSJetII-04 as the HEG for the conventional component and adapt it to generate the final state particles produced when classicalization occurs.  We only modify the treatment of particles with energy above $E_0 = 10^{xxx}$ eV.  We perform the shower simulations for assumed values of $M*$, $\mu$, $\kappa$ and $f_{K}$, then refine the values of these paramters to optimize the fit to the elongation rate and \rmsx\ above $10^{18.5}$ and the muon content observed in the ground signal of hybrid showers at $10^{19}$ eV.   

Given $M*$ and $\mu$, $f_{\rm cl}(E) \equiv \sigma_{\rm cl}(E)/\sigma_{\rm tot}(E)$ is the probability that the next interaction at energy $E$ produces a classicalized state.  Above a projectile energy $E_{0}$ we modify the standard simulation as follows: \\
1) Randomly assign the next interaction to be a classicalized or conventional one, according to the probability $f_{\rm cl}(E)$.\\
2)  Generate the next interaction point, using $\sigma_{\rm cl}(E)$ or $\sigma_{\rm HEG}(E)$ to compute the interaction length as appropriate.\\
3)  Generate the final state of the interaction using the classicalization or conventional HEG.  

Our classicalized event generator, CEG, is obtained as follows.  Choose an inelasticity cutoff, characteristic of central p-Air collisions in the chosen conventional HEG.  At each energy accept and reject events produced by the HEG according to whether their inelasticity is above or below the cutoff value, and find the mean multiplicity of the inelastic final states for that HEG: $<n_{\rm inel}>(E)$; define the multiplicity multiplier $f_{\rm mult}(E) \equiv N_{\rm cl}(E) / <n_{\rm inel}>(E) $.  To generate a classicalized final state:\\
a) Generate events at the given energy and for the given incident particle type with the conventional HEG, until an event with inelasticity above the cutoff is obtained.\\
b) Increase the multiplicity of the final state by randomly splitting a fraction $f_{\rm mult}(E)-1$ of the final particles into two particles with the given total energy; if $f_{\rm mult}(E) < 1$, combine by the same prescription.  (Because all but a tiny fraction of particles are highly relativistic, and detailed particle properties don't matter for the next stage of the interactions, the minute lack of energy-momentum and quantum number conservation is irrelevant.) 
c) Adjust the $K/\pi$ ratio of the final state to equal $r_{K}$ by randomly converting pions to kaons.

Conventional HEGs include a distribution of diffractive and inelastic final states as appropriate to p-Air (and more generally, for subsequent interactions, hadron-Air) collisions, reflecting both the intrinsic distribution found in $p-N$ collisions and the effect of the differing impact parameters of the p-nucleon collision and the profile of the nucleus.  As the primary energy increases, the classicalized state -- which initially is formed exclusively in central collisions -- is formed at progressively larger and larger impact parameters, with the conventional interactions occuring in more peripheral collisions.  This would result in some shift in the relative importance of diffractive and inelastic final states produced in the conventional collisions.  However the impact of this effect is of secondary importance and we need not incorporate this refinement in a first stage.  Thus the HEG is used as-is for the case the interaction is conventional.  


An unusual feature of the classicalization phenomenon is that there is a correlation between the depth of interaction and whether the classicalized state is formed.  Thus, the relationship between $S_{\mu}$ and \xmax\ in individual events is closer to that of a mixed composition, than in the pure proton  new-physics scenarios studied in \cite{afICRC13}.  This makes it more challenging to disentangle it from a mixed composition using the $S_{\mu}$ and \xmax\ correlations.  Nonetheless, the two scenarios have distinct features: \\
$\bullet$ In the classicalization scenario, the elongation rate, \rmsx\ and energy of the hadronic shower at ground evolve smoothly and monotonically with energy, whereas in a mixed composition scenario structure would be expected corresponding to the Peters cycle dropping out of lighter elements with different spectral normalizations. \\
$\bullet $ In a mixed composition scenario, one would expect that subtle structures in the spectrum corresponding to deviations from smooth composition evolution (certain mass-ranges that may be more or less dominant than obtained for a smooth dependence on ln<A>) would be reflected in correlated structures in the elongation rate and \rmsx\, whereas in the classicalization scenario the structure can only arise in the spectrum, as produced by source inhomogeneities, energy losses and transient sources; if seen at all, such spectral structures could be different for N and S hemisphere detectors.
\\
$\bullet $ Depending on $\mu$ and $M*$, in the classicalization scenario at highest energies the \xmax\ and \rmsx\ could become even smaller than in a pure Fe model. 
\\
$\bullet$ In the classicalization scenario the primaries are by Ocam's razor mostly protons, so that anisotropies of UHECR arrival directions should be consistent with production in in the local matter distribution, and deflections of protons by the Galactic and extragalactic magnetic fields.

In addition, there may be signatures at LHC which can be explored.

\section{Index of refraction at UHE}

An intriguing possibility is that due to QCD or a new interaction, the forward elastic scattering cross section becomes sufficiently large at UHE that hadrons propagating through the atmosphere experience an index of refraction greater than 1.  In this section we explore the implications of a sufficiently large index of refraction, and whether such effects could be large enough to have observational effects.  

From standard wave mechanics, the relationship between the forward scattering amplitude $f(0)$ and the index of refraction is
\be
n-1 = \frac{2 \pi \, n \, f(0) }{k^{2}},
\ee
where $k$ is the wavenumber and $n$ is the number density of scattering centers in the medium.  Using the optical theorem, 
$\sigma_{\rm tot} = \frac{4 \pi}{k} \, {\rm Im} f(0)$,
and taking $f(0) \approx {\rm Im} f(0)$, in the relativistic limit $k\approx E$ we have 
\be
\label{n-1}
n - 1 \approx
\frac{\rho \, \sigma_{\rm tot}}{s}, 
\ee
where $\rho$ is the mass density of the air and depending on whether the interaction is coherent or incoherent over the air nucleus $\sigma_{\rm tot}$ and $s$ refer to the p-Air or p-N cross section and $s$. 

$t-$channel exchange of a spin-2 particle of mass $\mu$ leads to $\sigma_{\rm el} \sim {s}$ at asymptotically high energy, with $\frac{d \, \sigma}{d \, t} ~ e^{-t/\mu^{2}}$.  For sufficiently small $\mu$, the interaction is so forward-peaked that $\sigma_{\rm el}$ could be much greater than $\sigma_{\rm inel}$ without being detected at the LHC, while being small enough at lower energies that it is hidden by conventional hadronic interactions.  
  
In a medium, however, Cerenkov radiation would occur for $\beta > n^{-1}$.  The radiated quanta are  
confined to a cone of opening angle
\be
\label{sinsqthetaC}
{\rm sin}(\theta_{C})^{2} = 1 - \frac{1}{n^{2} \beta^{2}} \approx 2 \frac{\rho \, \sigma_{\rm tot}}{s} - \frac{1}{\gamma^{2}}
\ee
and have a flat spectrum until cutoff at high energy.  This produces an energy loss per unit pathlength
\be
\label{Eloss}
\frac{d \, \gamma}{d \, x} \sim \alpha_{eff} \, m \, f(\gamma) {\rm sin}(\theta_{C})^{2},
\ee
with $f(\gamma)$ governed by the high energy cutoff of the hadronic Cerenkov radiation.  If the Cerenkov radiation were in the form of glueballs or other hadrons, and set in below a column depth of $\approx 50 {\rm g \, cm^{-2}}$ for $\gamma \sim 10^{9}$, the effect might mimic an enhanced inelastic cross section for proton primaries above $ 10^{18}$ eV and potentially account for the observed \xmax\ distribution.  

Charming as such an effect would be, it cannot be realized within QCD.  A lower limit on the $\gamma$ for QCD-Cerenkov radiation occurs can be obtained by  taking the cross section to increase as $\sim s$ starting at $10^{18.3} {\rm eV }$, where Auger has measured $\sigma_{pN} = 110 \pm $ mb \cite{augerXsecnPRL}.  Thus $\gamma \gtrsim 10^{14}$ is required to have QCD-Cerenkov radiation at a column depth of $50 {\rm g \, cm^{-2}}$.  For a proton of $10^{18.5}$ eV to be moving faster than the speed of light in the atmosphere at $\approx 50 {\rm g \, cm^{-2}}$ would require the cross section for the new spin-2 interaction to be a factor $\approx 8 \times 10^{4}$ times larger than the observed inelastic cross section at $\sqrt{s} = 57 $ TeV.  Such an interaction would make a 1 mb contribution to the elastic scattering cross section in a 200 GeV fixed-target experiment -- likely unobservably small.  Further study would be required whether the corresponding Cerenkov-like radiation could be hadronic; if not, it would not
have the desired effect.

A second interesting consequence of an index of refraction for hadrons is the suppression of the decay of $\pi^{0}$'s with $\gamma  >  1/ \sqrt{2 (n-1)}$.  Above this threshold, its energy is too small for it to decay into two photons of the same total momentum moving at $c$.  This is the same condition as for Cerenkov radiation, and hence would be realized for similar conditions to those discussed above.  Suppression of $\pi^{0}$ decay can explain the shortfall of muons in current HEGs, and combined with an evolving, mixed composition to account for the \xmax distribution, gives a phenomenologically viable model, as will be described elsewhere. 

\section{Acknowledgements}

GRF acknowledges with appreciation the very productive collaboration with J. D. Allen, which lead to these results.  She thanks G. Dvali for discussions about applying classicalization to QCD and his hospitality at Ludwig Maximillian University while this work was carried out, interesting correspondence with J. R. Cudell, and valuable discussions with colleagues the Pierre Auger Collaboration.  This research was supported in part by the National Science Foundation under grants NSF-PHY-1212538, NSF-PHY-0900631 and NSF-PHY-0970075.  


\end{document}